\documentclass[12pt]{article}
\usepackage{amssymb,amsfonts}
\usepackage{epsf,epsfig}
\begin{document}
\baselineskip 18pt

\title{Three-magnon bound states in exactly rung-dimerized spin ladders}
\author{P.~N.~Bibikov\\ \it Saint-Petersburg State University,\\
\it 198504 Ulyanovskaya 1, Petrodvorets, Saint-Petersburg, Russia}

\maketitle

\vskip5mm

\begin{abstract}
Three magnon bound state problem is studied within Bethe Ansatz for an exactly rung-dimerized spin ladder.
It is shown that contrary to the general three-magnon problem the solvability is less sensitive
to non integrability. All obtained wave functions are presented in explicit forms.
\end{abstract}

\maketitle

\section{Introduction}

During the last century a wide number of methods were developed for analysis of quantum {\it integrable} systems
\cite{1,2,3}. However almost all physically interesting models are non integrable. It is always assumed that in this
case all the analytical methods (which may be successfully applied for integrable models) should fail. In the present
paper we show that the situation is not so extremely hopeless. Treating the general model of exactly rung-dimerized
spin ladder \cite{4,5} we obtain general conditions for solvability of three-magnon bound state problem within Bethe
Ansatz. Surprisingly it turned out that these conditions are essentially weaker than the corresponding integrability ones.

Bethe Ansatz approach to three-magnon states in an exactly rung-dimerized spin ladder \cite{4} was previously
studied by the author \cite{5}. It was shown that the general problem is completely solvable only in five special
integrable cases. Solvability of a more special bound state problem at the first glance
also implies integrability. Indeed a bound state may be obtained \cite{1,2,3} from scattering ones by analytic
continuation of the corresponding wave numbers. However one may choose another way suggesting Bethe Ansatz for
bound states as a separate problem when only bounded exponents should be leaved. Surprisingly
this approach is less sensitive to non integrability.

\section{Spin ladder Hamiltonian}

Spin ladder Hamiltonian acts on an infinite tensor product of spaces related to the ladder rungs
(numerated here by an index $n$)
\begin{equation}
{\cal H}=\prod_n\otimes h_n.
\end{equation}
For each $n$ one has $h_n={\mathbb C}^4={\mathbb C}^2\otimes{\mathbb C}^2$ where the two ${\mathbb C}^2$-factors
are representation spaces for two triples of $S=1/2$ spin operators ${\bf S}_{j,n}$ ($j=1,2$) associated with the
$n$-th rung
\begin{equation}
{\bf S}^a_{j,n}{\bf S}^b_{j,n}=\frac{i}{2}\varepsilon_{abc}{\bf S}^c_{j,n},\quad [{\bf S}^a_{1,n},{\bf S}^b_{2,n}]=0,
\quad a,b,c=1,2,3.
\end{equation}
With respect to action of the total rung-spin operator
\begin{equation}
{\bf S}_n={\bf S}_{1,n}+{\bf S}_{2,n}.
\end{equation}
each space $h_n$ may be decomposed into three-dimensional rung-triplet and one-dimensional rung-singlet subspaces
\begin{equation}
h_n=h_n^t\oplus h_n^s,
\end{equation}

In the present paper we use the following representation for Hamiltonian of an {\it exactly} rung-dimerized
spin ladder \cite{5}
\begin{eqnarray}
\hat H&=&\sum_nJ_1Q_n+
J_2({\bf\Psi}_n\cdot{\bf\bar\Psi}_{n+1}+{\bf\bar\Psi}_n\cdot{\bf\Psi}_{n+1})+J_3Q_nQ_{n+1}\nonumber\\
&+&J_4{\bf S}_n\cdot{\bf S}_{n+1}+J_5({\bf S}_n\cdot{\bf S}_{n+1})^2.
\end{eqnarray}
Here
\begin{equation}
Q_n=\frac{1}{2}{\mathbf S}_n^2,
\end{equation}
is rung-triplet projector
\begin{equation}
Q_n|_{h_n^s}=0,\quad Q_n|_{h_n^t}=1.
\end{equation}
The two Hermitian conjugated ${\rm su}(2)$-covariant operator triples \cite{5}
\begin{eqnarray}
{\bf\Psi}_n&=&\frac{1}{2}({\bf S}_{1,n}-{\bf S}_{2,n})-i{[}{\bf
S}_{1,n}\times{\bf
S}_{2,n}{]},\nonumber\\
{\bf\bar\Psi}_n&=&\frac{1}{2}({\bf S}_{1,n}-{\bf
S}_{2,n})+i{[}{\bf S}_{1,n}\times{\bf S}_{2,n}{]},
\end{eqnarray}
additionally to
\begin{equation}
[{\bf S}_n^a,{\bf\Psi}_n^b]=i\varepsilon_{abc}{\bf\Psi}_n^c,\quad
[{\bf S}_n^a,{\bf\bar\Psi}_n^b]=i\varepsilon_{abc}{\bf\bar\Psi}_n^c,\quad a,b,c=1,2,3,
\end{equation}
satisfy relations
\begin{equation}
{\bf S}_n^a{\bf\Psi}_n^b=0,\quad{\bf\bar\Psi}_n^a{\bf S}_n^b=0.
\end{equation}
Both (9) and (10) may be proved from (2). According to (6), (9) and (10)
\begin{equation}
[Q_n,{\bf\bar\Psi}_n]={\bf\bar\Psi}_n,\quad[Q_n,{\bf\Psi}_n]=-{\bf\Psi}_n.
\end{equation}
So ${\bf\bar\Psi}_n$ and ${\bf\Psi}_n$ may be treated as (neither Bose no Fermi) creation-annihilation operators
for rung-triplets. The subspace $h_n^s$ is generated by a single vector $|0\rangle_n$ for which
\begin{equation}
{\bf\Psi}_n|0\rangle_n=0,\quad{\bf S}_n|0\rangle_n=0,\quad Q_n|0\rangle_n=0.
\end{equation}
We shall utilize the following two basises of the space $h_n^t$. The covariant basis
\begin{equation}
|1\rangle_n^a={\bf\bar\Psi}_n^a|0\rangle_n,\quad a=1,2,3,
\end{equation}
for which according to (9)
\begin{equation}
{\bf S}_n^a|1\rangle_n^b=i\varepsilon_{abc}|1\rangle_n^c,
\end{equation}
will be used for representation of three-magnon states with total spin 0 and 1.
The standard ${\bf S}_n^3$-eigenbasis
\begin{eqnarray}
{\mathbf S}^{\pm}_n|\tilde1\rangle_n^{\pm}&=&0,\quad{\mathbf S}^{\pm}_n|\tilde1\rangle_n^{\mp}=
\sqrt{2}|\tilde1\rangle_n^0,\quad
{\mathbf S}^{\pm}_n|\tilde1\rangle_n^0=
\sqrt{2}|\tilde1\rangle_n^{\pm},\nonumber\\
{\mathbf S}^3_n|\tilde1\rangle_n^{\pm}&=&\pm|\tilde1\rangle_n^{\pm},\quad {\mathbf S}^3_n|\tilde1\rangle_n^0=0,
\end{eqnarray}
will be used for representation of three-magnon states with total spin 2 and 3.
According to (9) one can suggest the following representation
\begin{equation}
|\tilde1\rangle_n^{\pm}=\mp{\bf\bar\Psi}_n^{\pm}|0\rangle_n,\quad|\tilde1\rangle_n^0=|1\rangle_n^3,
\end{equation}
where
\begin{equation}
{\bf\bar\Psi}_n^{\pm}\equiv\frac{1}{\sqrt{2}}\Big({\bf\bar\Psi}^1_n\pm
i{\bf\bar\Psi}^2_n\Big),\quad
{\bf\Psi}_n^{\pm}\equiv\frac{1}{\sqrt{2}}\Big({\bf\Psi}^1_n\pm
i{\bf\Psi}^2_n\Big).
\end{equation}
The formulas
\begin{equation}
{\bf\Psi}^a_n|1\rangle^b_n=\delta_{ab}|0\rangle_n,\quad a,b,=1,2,3,
\end{equation}
may be obtained from (13) by Hermitian conjugation.
Analogously for the operators ${\bf\Psi}^{\pm}_n$ one have
\begin{equation}
{\bf\Psi}^{\pm}_n|\tilde1\rangle^{\mp}_n=\pm|0\rangle_n,\quad{\bf\Psi}^{\pm}_n|\tilde1\rangle^{\pm}_n=
{\bf\Psi}^{\pm}_n|\tilde1\rangle^0_n=0.
\end{equation}

Eqs. (7) and (11) result in the commutation relation
\begin{equation}
[\hat H,\hat Q]=0,
\end{equation}
where
\begin{equation}
\hat Q=\sum_nQ_n,
\end{equation}
is (according to Eq. (7)) the number operator for rung-triplets. As it follows from (20)
an infinite tensor product of rung-singlets
\begin{equation}
|0\rangle_{r-d}=\prod_n\otimes|0\rangle_n,
\end{equation}
is an eigenvector for $\hat H$. It was already mentioned \cite{4,5} that for rather
big $J_1>0$ it is the ground state of the system. In this
(exactly rung-dimerized) case the physical Hilbert space is an infinite direct sum of multi-magnon sectors
\begin{equation}
{\cal H}^{phys}=\sum_{m=0}^{\infty}{\cal H}^m,\quad \hat Q|_{{\cal
H}^m}=m.
\end{equation}

Interpretation of the Hamiltonian (5) in this case is clear. The first term describes the chemical potential of an
excited rung-triplet. The second one corresponds to rung-triplets kinetic energy. We shall imply that $J_2\neq0$.
The last three terms describe a spin-dependent interaction between two neighboring rung-triplets.

\section{General properties of three-magnon wave functions}
We shall use the notation $|S,k\rangle$ for a three-magnon state with total spin $S$. At $S>0$ it will be supplied
by an upper index. For representation of a $S=1$ state $|1,k\rangle^a$ will be used the basis (13) (so in this case
$a=1,2,3$). At $S=2,3$ using the basis (16) we shall represent only the vectors $|S,k\rangle^S$ with $S^z=S$.
Namely we shall treat the following states \cite{5}
\begin{eqnarray}
|0,k\rangle&=&\epsilon_{abc}\sum_{m<n<p} {\rm
e}^{ik(m+n+p)/3}b_0(k,n-m,p-n)
\dots|1\rangle^a_m\dots|1\rangle^b_n\dots|1\rangle^c_p\dots,\nonumber\\
|1,k\rangle^a&=&\sum_{m<n<p} {\rm e}^{ik(m+n+p)/3}\Big[
b^{(1)}_1(k,n-m,p-n)
\dots|1\rangle^a_m\dots|1\rangle^b_n\dots|1\rangle^b_p\dots\nonumber\\
&+&b^{(2)}_1(k,n-m,p-n)\dots|1\rangle^b_m\dots|1\rangle^a_n\dots|1\rangle^b_p\dots\nonumber\\
&+&b^{(3)}_1(k,n-m,p-n)\dots|1\rangle^b_m\dots|1\rangle^b_n\dots|1\rangle^a_p\dots\Big],\nonumber\\
|2,k\rangle^2&=&\sum_{m<n<p} {\rm e}^{ik(m+n+p)/3}[
b^{(1)}_2(k,n-m,p-n)\dots|\tilde1\rangle^+_m\dots(|\tilde1\rangle^+_n\dots|\tilde1\rangle^0_p-
|\tilde1\rangle^0_n\dots|\tilde1\rangle^+_p)...\nonumber\\
&+&
b^{(2)}_2(k,n-m,p-n)\dots(|\tilde1\rangle^+_m\dots|\tilde1\rangle^0_n-
|\tilde1\rangle^0_m\dots|\tilde1\rangle^+_n)\dots|\tilde1\rangle^+_p\dots,\nonumber\\
|3,k\rangle^3&=&\sum_{m<n<p} {\rm e}^{ik(m+n+p)/3}b_3(k,n-m,p-n)
\dots|\tilde1\rangle^+_m\dots|\tilde1\rangle^+_n\dots|\tilde1\rangle^+_p\dots
\end{eqnarray}

All the (reduced to center mass) wave functions $b_j(k,m,n)$ ($j=0,1,2,3$) have a physical
sense only at $m,n>0$. However within the Bethe Ansatz approach \cite{3} they should be
continued into the two unphysical boundary regions $(m=0,n>0)$ and
$(m>0,n=0)$. At $m,n>1$ the corresponding ${\rm Schr\ddot odinger}$ equations have an identical form \cite{5}
\begin{eqnarray}
&6J_1b_j(k,m,n)+J_2[{\rm e}^{-ik/3}b_j(k,m+1,n)+{\rm
e}^{ik/3}b_j(k,m-1,n)+{\rm e}^{-ik/3}b_j(k,m-1,n+1)&\nonumber\\
&+{\rm e}^{ik/3}b_j(k,m+1,n-1)+{\rm e}^{-ik/3}b_j(k,m,n-1)+{\rm
e}^{ik/3}b_j(k,m,n+1)]=Eb_j(k,m,n),&
\end{eqnarray}
for all $j=0,1,2,3$. Requiring correctness of (25) in the boundary regions $(m=1,n>0)$ and $(m>0,n=1)$
one immediately obtains the following systems of Bethe conditions \cite{5}
\begin{eqnarray}
{\rm e}^{ik/3}b_0(k,0,n)+{\rm e}^{-ik/3}b_0(k,0,n+1)&=&2\Delta_1b_0(k,1,n),\nonumber\\
{\rm e}^{-ik/3}b_0(k,m,0)+{\rm e}^{ik/3}b_0(k,m+1,0)&=&2\Delta_1b_0(k,m,1),\\
{\rm e}^{ik/3}b^{(1)}_1(k,0,n)+{\rm e}^{-ik/3}b^{(1)}_1(k,0,n+1)&=&(\Delta_2+\Delta_1)b^{(1)}_1(k,1,n)+
(\Delta_2-\Delta_1)b^{(2)}_1(k,1,n),\nonumber\\
{\rm e}^{ik/3}b^{(2)}_1(k,0,n)+{\rm e}^{-ik/3}b^{(2)}_1(k,0,n+1)&=&
(\Delta_2+\Delta_1)b^{(2)}_1(k,1,n)+(\Delta_2-\Delta_1)b^{(1)}_1(k,1,n),\nonumber\\
{\rm e}^{-ik/3}b^{(2)}_1(k,m,0)+{\rm e}^{ik/3}b^{(2)}_1(k,m+1,0)&=&
(\Delta_2+\Delta_1)b^{(2)}_1(k,m,1)+(\Delta_2-\Delta_1)b^{(3)}_1(k,m,1),\nonumber\\
{\rm e}^{-ik/3}b^{(3)}_1(k,m,0)+{\rm e}^{ik/3}b^{(3)}_1(k,m+1,0)&=&
(\Delta_2+\Delta_1)b^{(3)}_1(k,m,1)+(\Delta_2-\Delta_1)b^{(2)}_1(k,m,1),\nonumber\\
{\rm e}^{ik/3}b^{(3)}_1(k,0,n)+{\rm e}^{-ik/3}b^{(3)}_1(k,0,n+1)&=&
\frac{2}{3}(\Delta_0-\Delta_2)[b^{(1)}_1(k,1,n)+b^{(2)}_1(k,1,n)]\nonumber\\
&+&2\Delta_0b^{(3)}_1(k,1,n),\nonumber\\
{\rm e}^{-ik/3}b^{(1)}_1(k,m,0)+{\rm e}^{ik/3}b^{(1)}_1(k,m+1,0)&=&
\frac{2}{3}(\Delta_0-\Delta_2)[b^{(2)}_1(k,m,1)+b^{(3)}_1(k,m,1)]\nonumber\\
&+&2\Delta_0b^{(1)}_1(k,m,1),\\
{\rm e}^{ik/3}b_2^{(2)}(k,0,n)+{\rm e}^{-ik/3}b_2^{(2)}(k,0,n+1)&=&
2\Delta_1b_2^{(2)}(k,1,n)+(\Delta_2-\Delta_1)b_2^{(1)}(k,1,n),\nonumber\\
{\rm e}^{ik/3}b_2^{(1)}(k,0,n)+{\rm e}^{-ik/3}b_2^{(1)}(k,0,n+1)&=&2\Delta_2b_2^{(1)}(k,1,n),\nonumber\\
{\rm e}^{-ik/3}b_2^{(1)}(k,m,0)+{\rm e}^{ik/3}b_2^{(1)}(k,m+1,0)&=&
2\Delta_1b_2^{(1)}(k,m,1)+(\Delta_2-\Delta_1)b_2^{(2)}(k,m,1),\nonumber\\
{\rm e}^{-ik/3}b_2^{(2)}(k,m,0)+{\rm e}^{ik/3}b_2^{(2)}(k,m+1,0)&=&2\Delta_2b_2^{(2)}(k,m,1),
\end{eqnarray}
where
\begin{eqnarray}
\Delta_0=\frac{J_3-2J_4+4J_5}{2J_2},\quad
\Delta_1=\frac{J_3-J_4+J_5}{2J_2},\quad
\Delta_2=\frac{J_3+J_4+J_5}{2J_2}.
\end{eqnarray}
The system related to $b_3(k,m,n)$ has the form identical
identical to Eq. (26) however with $\Delta_1$ replaced on $\Delta_2$.

The systems (26)-(28) are invariant under the set of corresponding duality
transformations
\begin{eqnarray}
{\cal D}(b_{0,3}(k,m,n))&=&\bar b_{0,3}(k,n,m),\quad
{\cal D}(b^{(j)}_1(k,m,n))=\bar b^{(4-j)}_1(k,n,m),\nonumber\\
{\cal D}(b^{(j)}_2(k,m,n))&=&\bar b^{(3-j)}_2(k,n,m).
\end{eqnarray}

Since in the present paper we are interesting only in bound states
we imply a normalization condition
\begin{equation}
\sum_{m,n>0}|b_j(k,m,n)|^2<\infty,\quad j=0,1,2,3.
\end{equation}

One may ask a question in what interval lies $k$. On the one hand there should be
\begin{equation}
0\leq k<2\pi,
\end{equation}
on the other a substitution $k+2\pi$ evidently changes the exponential factors in (24). In order to clarify the
situation we notice that this substitution together with simultaneous multiplication of the wave function on the
factor ${\rm e}^{2\pi i(m-n)/3}$ retains invariant all the states in (24)

\section{Bound states at $\bf S=0$ ($\bf S=3$)}

The following substitution
\begin{equation}
b_0(k,m,n)=z_+^mz_-^n,
\end{equation}
solves Eq. (25) resulting in a dispersion
\begin{equation}
E(k)=6J_1+J_2\Big[{\rm e}^{ik/3}\Big(\frac{z_+}{z_-}+\frac{1}{z_+}+z_-\Big)
+{\rm e}^{-ik/3}\Big(\frac{z_-}{z_+}+\frac{1}{z_-}+z_+\Big)\Big].
\end{equation}
A substitution of (33) into Eq. (26) gives
\begin{equation}
2\Delta_1z_{\pm}={\rm e}^{\pm ik/3}+{\rm e}^{\mp ik/3}z_{\mp}.
\end{equation}
From Eq. (35) readily follows
\begin{equation}
2\Delta_1(z_--\bar z_+)={\rm
e}^{ik/3}(z_+-\bar z_-).
\end{equation}
In the general case
\begin{equation}
2|\Delta_1|\neq1,
\end{equation}
Eq. (36) results in an autoduality (with respect to (30)) condition
\begin{equation}
\bar z_-=z_+\equiv z,
\end{equation}
and the system (35) reduces to an equation
\begin{equation}
G_1=0,
\end{equation}
where
\begin{equation}
G_j\equiv2\Delta_jz-{\rm e}^{ik/3}-{\rm
e}^{-ik/3}\bar z.
\end{equation}
Eqs. (39), (40) result in
\begin{equation}
z=\frac{2\Delta_1{\rm e}^{ik/3}+{\rm e}^{-2ik/3}}{2\Delta_1^2-1}.
\end{equation}

A substitution of (38) and (41) into (34) gives
\begin{equation}
E(k)=6J_1+\frac{2J_2}{4\Delta_1^2-1}(8\Delta_1^3+\cos{k}).
\end{equation}

Condition (31) is satisfied only if $|z|<1$ or according to (41)
\begin{equation}
\Delta_1\cos{k}<(4\Delta_1^2-3)\Delta^2_1.
\end{equation}

The non-autodual states at $2|\Delta_1|=1$ will be studied in a separate paper. Here we shall treat
only autodual wave functions.

As it was mentioned above all calculations for $S=3$ may be performed in the similar manner after the
replacement $\Delta_1$ on $\Delta_2$.

\section{Bound states at $\bf S=1$}

Treating an autodual substitution
\begin{eqnarray}
b^{(j)}_1(k,m,n)&=&B^{(j)}z^m\bar z^n,\\
B^{(3)}&=&\bar B^{(1)},
\end{eqnarray}
one gets from Eq. (27)
\begin{eqnarray}
\left(\begin{array}{ccc}
G_1+G_2&G_2-G_1&0\\
G_2-G_1&G_1+G_2&0\\
G_0-G_2&G_0-G_2&3G_0
\end{array}\right)\left(\begin{array}{c}
B^{(1)}\\
B^{(2)}\\
B^{(3)}\end{array}\right)=0.
\end{eqnarray}

Since determinant of the matrix in Eq. (46) is proportional to
$G_0G_1G_2$ the system (46) is solvable in three cases
\begin{eqnarray}
G_j=0,\quad j=0,1,2.
\end{eqnarray}
The corresponding solutions in general do not satisfy Eq. (45). However this problem may be solved if the system (46)
has two different solutions. Taking into account that the latter is possible only if $\Delta_j=\Delta_l$ for $j\neq l$
we conclude that a three-magnon bound state related to the wave function exists at three cases
\begin{eqnarray}
\Delta_0&=&\Delta_1,\quad B=[1,-1,1],\\
\Delta_0&=&\Delta_2,\quad B=[1,1,1],\\
\Delta_1&=&\Delta_2,\quad B=[1,-4,1],
\end{eqnarray}
where each $B\equiv [B^{(1)},B^{(2)},B^{(3)}]$ represents the corresponding autodual solution of Eq. (46).

\section{Bound states at $\bf S=2$}

An autodual substitution
\begin{eqnarray}
b^{(j)}_2(k,m,n)&=&C^{(j)}z^m\bar z^n,\\
C^{(2)}&=&\bar C^{(1)},
\end{eqnarray}
results in
\begin{equation}
\left(\begin{array}{cc}
G_2&0\\
G_2-G_1&2G_1\\
\end{array}\right)\left(\begin{array}{c}
C^{(1)}\\
C^{(2)}
\end{array}\right)=0.
\end{equation}
The system (53) is solvable under one of the conditions (47)
taken for $j=1,2$. However autodual solutions $[1,1]$ and $[i,-i]$ exist only at $\Delta_2=\Delta_1$.

\section{Summary and discussion}

As it was previously shown in Ref. 5 the system related to Hamiltonian (5) is integrable only in five special cases
\begin{eqnarray}
\Delta_0&=&\Delta_1=\Delta_2,\\
\Delta_0&=&\Delta_2=\pm1,\quad\Delta_1=0,\\
\Delta_0&=&\Delta_2=0,\quad\Delta_1=\pm1,\\
\Delta_1&=&\Delta_2=\pm\frac{3}{2},\quad\Delta_0=0,\\
\Delta_1&=&\Delta_2=0,\quad\Delta_0=\pm\frac{3}{2}.
\end{eqnarray}
We see that restrictions (48)-(50) are essentially weaker than (54)-(58). So the bound states of Bethe form exist even
for a rather big number of non integrable models.

In order to clarify a relevance of conditions (48)-(50) to physical applications we use an equivalent representation of
the Hamiltonian (5) \cite{5}
\begin{equation}
H_{n,n+1}=J_rH^r_{n,n+1}+J_lH^l_{n,n+1}
+J_dH^d_{n,n+1}+J_{rr}H^{rr}_{n,n+1}
+J_{ll}H^{ll}_{n,n+1}+J_{dd}H^{dd}_{n,n+1},
\end{equation}
where
\begin{eqnarray}
H^r_{n,n+1}&=&\frac{1}{2}({\bf S}_{1,n}\cdot{\bf S}_{2,n}+{\bf
S}_{1,n+1}\cdot{\bf S}_{2,n+1}),\quad H^l_{n,n+1}={\bf
S}_{1,n}\cdot{\bf S}_{1,n+1}+{\bf S}_{2,n}\cdot{\bf
S}_{2,n+1},\nonumber\\
H^d_{n,n+1}&=&{\bf S}_{1,n}\cdot{\bf S}_{2,n+1}+{\bf
S}_{2,n}\cdot{\bf S}_{1,n+1},\quad H^{rr}_{n,n+1}=({\bf
S}_{1,n}\cdot{\bf S}_{2,n})({\bf S}_{1,n+1}\cdot{\bf
S}_{2,n+1}),\nonumber\\
H^{ll}_{n,n+1}&=&({\bf S}_{1,n}\cdot{\bf S}_{1,n+1})({\bf
S}_{2,n}\cdot{\bf S}_{2,n+1}),\quad H^{dd}_{n,n+1}=({\bf
S}_{1,n}\cdot{\bf S}_{2,n+1})({\bf S}_{2,n}\cdot{\bf S}_{1,n+1}).
\end{eqnarray}
A correspondence between the coupling constants of (5) and (59) is given by Eq. (20) of Ref. 5.
We imply also the condition \cite{4,5}
\begin{equation}
J_{ll}-J_{dd}=4(J_l-J_d),
\end{equation}
which guarantees the commutation relation (20).
According to Eq. (32) of Ref. 5
\begin{eqnarray}
\Delta_0=\Delta_1&\Leftrightarrow&12J_l-8J_d-5J_{ll}=0,\\
\Delta_0=\Delta_2&\Leftrightarrow&4J_l-J_{ll}=0,\\
\Delta_1=\Delta_2&\Leftrightarrow&4J_d+J_{ll}=0.
\end{eqnarray}

It is assumed \cite{6} that for physically relevant models there should be
\begin{equation}
|J_{rr}|,|J_{ll}|,|J_{dd}|<|J_l|,
\end{equation}
so Eq. (61) may be satisfied only for highly frustrated ladders with
\begin{equation}
J_d\approx J_l.
\end{equation}
In this context the case (62) seems to be physically more reliable.

The author is very grateful to P. P. Kulish for helpful discussions.

\end{document}